\documentclass{aa}

\usepackage{psfig,amsmath,txfonts,amssymb,graphicx}

\newcommand{\nc}{\newcommand}
\nc{\Porb}{$P_{\rm orb}$\,}
\nc{\Teff}{$T_{\rm eff}$\,}
\nc{\logg}{log\,$g$\,}
\nc{\kms}{\,${\rm km\,s}^{-1}$\,}
\nc{\Msun}{$M_{\odot}\ $}
\nc{\Mcz}{$M_{CZ}\ $}
\nc{\vsini}{$v \sin i$}
\nc{\vrad}{$v_{\rm rad}$}
\nc{\ALi}{A$_{\rm Li}$\,}
\nc{\Ali}{A$_{\rm Li}$\,}

\def\ub{\hbox{$U\!-\!B$}}               
\def\bv{\hbox{$B\!-\!V$}}

\begin{document}

\title{Lithium and magnetic fields in giants.\\
HD 232\,862 : a magnetic and lithium-rich giant star.
\thanks{Based on spectropolarimetric observations obtained at the Canada-France-Hawaii 
Telescope (CFHT, operated by 
the National Research Council of Canada, the Institut National des 
Sciences de l'Univers of the Centre National de la Recherche 
Scientifique of France, and the University of Hawaii), and at the T\'elescope Bernard Lyot (TBL at 
Observatoire du Pic du Midi, CNRS and Universit\'e de Toulouse, France).
}
}

\author{   A. L\`ebre \inst{1}
\and             A. Palacios \inst{1}
\and          J. D. do Nascimento Jr \inst{2}
\and             R. Konstantinova-Antova \inst{3}
\and             D. Kolev \inst{3}
\and             M. Auri\`ere \inst{4}
\and             P. de Laverny \inst{5} 
\and             J. R. De Medeiros \inst{2} 
}
\offprints{A. L\`ebre\\
 lebre@graal.univ-montp2.fr}

\institute{ Groupe de Recherche en Astronomie et Astrophysique du
Languedoc, UMR5024, Universit\'e Montpellier II, CNRS, 
      Place E. Bataillon, 34095 Montpellier, France
                 \and
Departamento de F\'{\i}sica,
      Universidade Federal do Rio
      Grande do Norte, 59072-970
      Natal,  RN, Brazil  
       \and
       Institute of Astronomy, Bulgarian Academy of Sciences, 
       72 Tsarigradsko shose, BG-1784 Sofia, Bulgaria
       \and
       Laboratoire d'Astrophysique de Toulouse-Tarbes, Universit\'e de Toulouse, 
       CNRS, Observatoire Midi-Pyr\'en\'ees, 57 avenue d'Azereix, 65008 Tarbes, France
                 \and
      Cassiop\'ee UMR 6202,
      Universit\'e de Nice Sophia Antipolis, CNRS,  
      Observatoire de la C\^ote d'Azur,
      BP 4229, 06304 Nice, France}
\date{Received  March 2009 ; accepted June 2009}
\authorrunning{L\`ebre et al.,}
\titlerunning{HD 232\,862 : a magnetic \& Li-rich giant star}

\abstract
{}
{We report the detection of an unusually high lithium content in HD 232\,862, a field giant classified as a G8II star, 
and hosting a magnetic field. }
{With the spectropolarimeters ESPaDOnS at CFHT and NARVAL at TBL, we have collected 
high resolution and high signal-to-noise spectra of three giants : HD 232\,862, KU Peg and HD 21\,018.  
From spectral synthesis  we have inferred  stellar parameters and measured  lithium abundances that we have 
compared to predictions from evolutionary models.\\ 
We have also analysed Stokes V signatures, looking for a magnetic 
field on these giants.}  
{HD 232\,862, presents a very high abundance of lithium (A$_{\rm Li}$\,= 2.45 $\pm$ 0.25 dex), far in excess of the 
theoretically value expected at this spectral type and for this luminosity 
class (i.e, G8II). The evolutionary stage of HD 232\,862 has been precised, and it suggests a mass in the lower part of the 
[1.0 M$_\odot$, 3.5 M$_\odot$] mass interval, likely 1.5 to 2.0 \Msun, at the bottom of the Red Giant Branch. 
Besides, a time variable Stokes V signature has been detected in the data of HD 232\,862 and KU Peg, pointing 
to the presence of a magnetic field at the surface of these two rapidly rotating active stars. }
{}


\maketitle

\section{Introduction}

Lithium-rich giant stars represent a real puzzle for stellar astrophysics. 
They are essentially late G or K type stars with 
Li abundance far in excess with respect to the values predicted by standard theory. 
In this latter, the fragile element lithium is expected to be 
destroyed in all but the outermost layers of a Main Sequence (MS) star. 
On the ascent of the Red Giant Branch (RGB), the convective envelope deepens (in mass) during the so-called first dredge-up episode. 
The preserved lithium of the envelope is then diluted into deeper Li free regions, 
reducing its surface abundance by a large factor (Iben 1965,1966a,b,1991).\\

The discovery by Wallerstein \& Sneden (1982) of the first K giant with a nearly cosmic Li abundance 
(A$_{\rm Li}$ = 3.0 dex) has challenged the standard scenario of Li dilution along stellar evolution. 
Since then, an increasing number of such stars has appeared in the literature (e.g., 
Brown et al., 1989; Gratton \& D'Antona  1989; de la Reza et al., 1996; Kraft et al., 1999 ; Reddy \& Lambert 2005; 
Monaco \& Bonifacio 2008 ; Roederer et al., 2008). 
Although most of these Li-rich giants were found to be slow rotators (De Medeiros et al., 1996a and 1996b), the report of moderate
to fast rotating Li-rich giants (i.e., with \vsini~ $>$ 8 \kms; Fekel \& Balachandra 1993 ; Fekel et al., 1996; 
Drake et al., 2002; Reddy et al., 2002; de Laverny et al., 2003), points to a link between rotation and 
high-Li content (de Laverny et al., 2003).\\

The Li-rich phenomenon has been observed in very few (less than 5 \%) red giant
  stars in the field, in globular and open clusters (Pilachowski et al., 2000;
  Charbonnel \& Balachandran 2000; Hill \& Pasquini 2000), and it has thus  been suggested to be
  associated with a very short-lived episode of surface Li
  enrichment. The root cause for this enrichment has been
hypothesized following two main directions.  
The first one concerns
external processes, like the contamination of the upper layers of the
star by the debris of nova ejecta or by the accretion of material in
the form of planets or a brown dwarf (Alexander, 1967; Brown et
al. 1989; Gratton and D'Antona 1989; Siess and Livio 1999; Denissenkov
and Weiss 2000; Reddy et al., 2002). 
The second one is related to internal processes, either preserving the initial Li from being completly 
diluted during the first dredge-up, or leading to the production of fresh 
lithium via the Cameron \& Fowler (1971) mechanism (Fekel \& Balachandran 1993; de la Reza et al., 1996; Sackmann \& 
Boothroyd 1999; Charbonnel \& Balachandran, 2000; Palacios et al., 2001).\\ 
This last scenario appears to be the most favoured by the additional abundance 
constraints available for the Li-rich giants concerning boron and  beryllium, that are not enhanced 
contrary to what would be expected if Li enrichment was due to planet or brown dwarfs engulfment (see 
Melo et al., 2005 ; Castilho et al., 1999,  de la Reza,
  2006).  Charbonnel \& Balachandran (2000) have identified the
  evolutionary status of the low-mass Li-rich giants, and after
  discarding those undergoing Li dilution via the first dredge-up
  episode, these peculiar stars lie at the RGB bump. They all present
  normal carbon isotopic ratio, contrary to stars beyond the bump
  which usually (it is the case for 96\% of them) present
  over-depleted $^{12}{\rm C}$ and low $^{12}{\rm C}/^{13}{\rm
    C}$. These authors suggest that the small statistics of Li-rich objects 
  and their precise location at the bump indicate a short-lived
  production episode occuring at the point where the mean molecular
  weight barrier separating the convective envelope from the hydrogen
  burning shell is erased. \\

Another puzzling aspect concerns the class of luminosity of these Li--rich stars, essentially giants 
of luminosity class III. To date, there is no clear register of Li-rich stars among the field 
subgiants, namely the luminosity class IV stars (De Medeiros et al., 1997; L\`ebre et al., 1999; 
Randich et al., 1999). In open clusters however, Canto Martins et al., (2006) have reported the 
detection of an exceptionally high Li abundance for a binary subgiant in M67. They have explored 
two hypotheses to explain such a high Li content : binarity and related tidal effects on one hand, 
and atomic diffusion on another hand. Among the field bright giant stars, namely the luminosity class II stars, 
there is also no clear register of Li-rich objects (Luck \& Wepfer, 1995 ; 
L\`ebre et al., 2006). \\

In this paper we report on spectropolarimetric observations of three late--G type field giants : 
HD 232\,862, KU~Peg (= HD 218\,153) and  HD 21\,018.  
Sample stars and observational data are described in Section~2.  
Our initial aim was to investigate lithium abundance in these objects. The methodology we used 
to derive stellar parameters and to measure Li abundance is described in Section~3. However, the use of 
instruments such as spectropolarimeters also enabled us to explore the presence of surface magnetic field. 
In Section 4, we expose the first direct detections of a magnetic field at the surfaces of HD 232\,862 and  
KU~Peg. In Section 5, lithium contents are discussed, considering evolutionary status of the objects. 
Conclusions are presented in Section 6.

\section{Sample stars and observational data}

\subsection{HD 232\,862}

From the SIMBAD database, HD 232\,862 is presented as a star with a G8 spectral type and a class of luminosity II,  
namely a bright giant star, with a mass range from about 2.5 to 9 \Msun. 
Very few publications have been devoted to this object, and among them several deal with its binary status 
(see further, Section 2.3). Among the few available measurements, one can only find IUE spectra, and photometric and CORAVEL 
data. Unfortunately no HIPPARCOS parallax is available for this star. \\

Photometric observations of HD 232\,862 were carried out with the computer-controlled single-channel 
electrophotometer 
attached at the 60~cm Cassegrain telescope at the Belogradchik Observatory (Bulgaria), during 
5 nights (Feb. 21 \& 22, 1998 and Jan. 5,6 \& 7, 1999). The equipment is described by Antov and 
Konstantinova-Antova (1995). From these 5 nights measurements, the following mean $UBV$ estimations 
were obtained : \bv = 0.87 $\pm$ 0.02 ; \ub = 0.41 $\pm$ 0.05 ; $V$ = 9.60 $\pm$ 0.04. Because 
the two components of this binary system (see Section 2.3) were likely in the diaphragm of the electrophotometer, 
the  \ub~index is bluer than expected for a G8III or a G8II star (\ub = 0.69, for  a G8III star). 
From these photometric data, we have estimated the effective temperature from the \Teff - \bv ~ 
calibration by Flower (1996). As a very first guess to compute synthetic spectra (see Section~3), 
we have found \Teff = 5\,120 $\pm$ 40~K.\\
 According to Bell \& Gustafsson (1989), this value is in good 
agreement with a G8 spectral type (and a \bv~ index of 0.87), but with a class of luminosity III rather than II.\\

From CORAVEL measurements (De Medeiros \& Mayor, 1999), HD 232\,862 presents a rather constant radial velocity and a 
rotational velocity (\vsini) of 20.6 \kms, a value considered as very high regarding its spectral type and its 
luminosity class (G8II). Indeed, as reported by De Medeiros et al., (1996) the mean rotational velocity of late-G bright giants ranges 
from 3.2 \kms at G5II to 2.1 \kms at G9II. Hence we can not exclude a mis-classification of this object among the class 
of luminosity II ; the estimation of its mass (see Sect.~5) seems also to confirm this fact.\\ 

HD 232\,862 is also reported as a X-ray source in the ROSAT data basis. It 
shows interesting features in the ultraviolet from IUE observations, pointing to a coronal and 
chromospheric activity. Indeed, on the basis of UV line fluxes measured from IUE SWP 42306, a 240 mn exposure 
at low dispersion taken on August 22, 1991, one can observe a large number of strong lines that originate from 
the high ionization states of N, Si and C. These features have also been reported in the spectra of chromospherically active giants (e.g., KU Peg, see below).

\subsection{KU Peg and HD 21\,018}

After the detection of important Li feature and magnetic filed
  signature in the spectrum of HD 232\,862, observations on two other
field giants have been collected with similar instrumentation
(spectropolarimetry, see below) for comparative purposes.   Two
  rapidly rotating stars have been selected from the De Mederiros and
  Mayor (1999) catalogue. KU Peg and HD 21\,018 present spectral
  types, and hence expected stellar parameters similar to HD
  232\,862. KU Peg has been selected for stellar activity aspect and 
  surface magnetic field search, while HD 21\,018 has been observed
  for Li content and evolutionary aspects. \\

The G8II bright giant KU Peg (= HD 218\,153) is one 
of the most chromospherically active stars, only the peculiar FK Comae star being significantly more active (De 
Medeiros et al., 1992). It is a rapidly rotating star (\vsini~= 27.6 \kms) and Strassmeier \& Hall (1988) have found 
its photometric period ($\sim$ 24 days) to be interpreted as the stellar rotation period. 
De Medeiros and L\`ebre (1992) have produced a spectroscopic study of KU Peg in the lithium line (670.7~nm) region.  
They  have reported no sign of significant LiI feature in this rapidly rotating giant. This fact is  in good 
agreement with the evolutionary status of KU Peg and the expected on-going Li dilution occurring through 
the first dredge-up episode at the bottom of the RGB. \\
More recently, Weber et al., (2005), by confronting Doppler images of KU Peg to differential-rotation 
models (Kitchatinov \& R\"udiger, 2004), claimed that this rapidly rotating active star could host a weak 
magnetic field (less than 100 G). To date however, direct detection of such a magnetic field in KU Peg is still lacking. \\

The G5III giant HD 21\,018 (= HR 1\,023) is known as a rapidly rotating Li-rich giant (Hartoog, 1978 ; 
Barrado y Navascues et al., 1998). According to Charbonnel and Balachandran (2000) and  Dennissenkov et al., (2006), 
this high Li content could be explained by the evolutionary status of the star : an intermediate-mass ($>$ 2.5 \Msun) star 
on its way from Main Sequence to RGB. \\

Table~1 presents the list of the three giants involved into this work, together with useful parameters.

\begin{table}
\caption{List of the program stars. V magnitude (V) and spectral type are taken from SIMBAD database. 
\vsini~ and \vrad ~values (in \kms), and \bv~ are taken from the catalogue of De Medeiros \& Mayor (1999). 
}
\begin{flushleft}
\begin{tabular}{ l l l l l l }
\noalign{\smallskip}
\hline
\hline
star  & V & Spectral  & \bv & \vsini & \vrad \\
 &  &type & & (\kms) & (\kms)\\
\hline

HD 232\,862 & 9.46 & G8II &  0.87 & 20.6 & - 1.80  \\
HD 21\,018  & 6.40 & G5III& 0.86 & 22.7 &   9.16\\
KU Peg  & 7.66 & G8II & 1.12 & 27.1 & - 80.49 \\

\hline
\end{tabular}
   \end{flushleft}  
\end{table}

\subsection{Spectropolarimetric observations}

Spectropolarimetric observations of HD 232\,862 have been collected in December 2006 with the new generation 
of spectropolarimeter, ESPaDOnS, installed at the CFH Telescope (CFHT, a 4m class telescope) in Mauna Kea (Hawaii, USA). 
The ESPaDOnS instrument consists in a cross-dispersion echelle spectrograph and a polarimetric module 
(Donati 2004; Donati et al., 2006). Spectropolarimetric data on the two  
other giants, KU Peg and HD 21\,018 , have been collected in September 2007 with 
the ESpaDOnS'twin brother, NARVAL (Auri\`ere, 2003), installed at T\'elescope Bernard Lyot 
(TBL, a 2m class telescope) at Pic du Midi (France). \\

For all observations (with ESPaDOnS and with NARVAL), the polarimetric mode has been used. It enables 
a very large coverage of the spectral region (from 375 nm to 1\,050 nm) and provides a spectral resolution of 
about 65\,000. The observations were all carried out in the circular polarization mode in order to collect 
simultaneous informations in Stokes V and I parameters.  One observation consists of a sequence of four consecutive 
exposures taken with different waveplate configurations. Data reduction (including bias and Flat 
Field correction, order extraction, wavelength calibration and normalization to the continuum) was achieved 
with the Libre-ESpRIT routines (adapted from the software package ESpRIT, see Donati et al., 1997). This software was 
made available on line at CFHT and at TBL during the observing runs, as well as other tools (see Sect. 4) . \\
 
HD 232\,862 has been observed each night, from the 7th to
the 10th of December 2006. In the 2007 September run, with
NARVAL, KU Peg has been observed during 
two consecutive nights, while HD 21\,018
has been observed only once because no magnetic detection was 
found from its Stokes V profile (see Sect.~4). The log of all our
spectropolarimetric observations (ESPaDOnS and NARVAL) is presented
in Table~2. \\

\begin{table}
\caption{Log of our spectropolarimetric ESPaDOnS and NARVAL observations.
Julian date (+ 2 454 000.000) is always considered at mid-exposure (i.e., at the beginning of the third 
subexposure). Total exposure time of a complete V sequence (i.e., cumulating 4 subexposures) is given (in sec.).  
Maximum signal-to-noise ratio (S/N) - on I spectrum - is reported (per velocity bin 
of 2.6 \kms, near 730 nm). 
}
\begin{flushleft}
ESPaDOnS--CFHT Observations -- December 2006 :\\
\smallskip
\begin{tabular}{l l l c l }
\noalign{\smallskip}
\hline
\hline
star  & date & Julian    & exposure & S/N  \\
  & dd/mm/yyyy &  date   & time &    \\

\hline

HD 232\,862  & 07/12/2006 & 77.937 &  3600 &  307 \\
HD 232\,862  & 08/12/2006 & 78.751 &  2800 &  254 \\
HD 232\,862  & 09/12/2006 & 79.826 &  2000 &  157  \\
HD 232\,862  & 10/12/2006 & 80.818 &  2400 &  278 \\
\hline
\end{tabular}
\medskip

NARVAL--TBL Observations -- September 2007 : \\
\smallskip

\begin{tabular}{ l l l c l }
\noalign{\smallskip}
\hline
\hline
star  & date & Julian    & exposure & S/N   \\
  & dd/mm/yyyy &  date   &  time &    \\
\hline

KU Peg  &  04/09/2007 & 348.516 & 1200 &  320\\
KU Peg  &  05/09/2007 & 349.502 & 1200 &  344\\
HD 21\,018   &  05/09/2007 & 349.633 &  800 &  391\\
\hline
\end{tabular}
   \end{flushleft}  
\end{table}

\subsection{Binary status of HD 232\,862}
HD 232\,862 (= cou2357) has been resolved as a visually tight binary by Couteau (1988). 
This star was measured 4 times in the interval 1988.04 - 1997.10 : twice by Couteau in 1988 and 1997 
(Couteau 1988, and Gilli et al., 1997) and twice by Heintz in 1990 and 1998 (Heintz 1990 and Heintz 1998). 
These authors measured it as being composed of two tenth magnitude 
stars separated by about 0.75-0.85 arcec. No significant changes in the relative positions of the 
components were observed during the measurements interval, which suggests a very long period. \\

The components of COU2357 were easily separated on the guiding camera of ESPaDOnS during our 
subarcsec seeing observations at CFHT. We thus put the giant star in the 1.58" hole and guided visually on 
the other star whose image was completely out of the hole. We roughly estimated the distance between the 
two components to be less than 2 arcsec which may suggest a much larger displacement of the components 
during the last decade than in the previous one (possible elliptical orbit). The giant star was also 
significantly brighter than its companion on the guiding CCD camera. This may suggest a secondary star 
hotter than the giant star. Because of the rather small separation between the components of this binary 
we consider that the CORAVEL observations (and more generally 2m class telescope observations)  
involved both stars, as well as previous photometric observations. 
Because of the certainly very long period of the binary and a low associated differential radial 
velocity, CORAVEL could not detect the binary nature of HD 232\,862. 
Our ESPaDOnS observations thus appear as the first spectra of the giant 
star alone. 

\section{Stellar parameters and Lithium abundance}

We made use of synthetic spectra to derive stellar parameters and lithium abundances of our 
three program stars. The MARCS stellar atmosphere models (Gustafsson et al., 2008) 
and the spectral synthesis tools TURBOSPECTRUM (described in Alvarez \& Plez 1998) have been used.  
Solar abundances have been taken from Grevesse and Sauval (1998). 
Using a first guess on stellar model parameters (\Teff = 5000 K ; \logg =
2.0 and [Fe/H] = 0.0) and a microturbulence velocity always set to
2.0 \kms, we have computed synthetic spectra. We used a grid of MARCS models 
presenting a step in \Teff of 250~K and a step in \logg of 0.5~dex, while specific 
abundances as well as metallicity ([Fe/H]) could be adjusted precisely. 
We have convolved the resulting synthetic spectra by a gaussian profile, in order to reproduce the instrumental
profile of ESPaDOnS and NARVAL, and by a rotational profile so as to
take into account the \vsini~ of each object (see Tab. 1).\\ 

To derive stellar parameters and to measure lithium abundances, under Local Thermodynamics Equilibrium (LTE) hypothesis, 
the spectral synthesis has been performed through the Balmer lines (H$\alpha$ and  H$\beta$) and the Li line (around 6710 \AA) regions. 
For this Li region, we made used of the specific atomic 
line list described in Canto Martins et al., (2006), initially issued from VALD (Kupka et al., 1999, Ryabchikova et al., 1999) 
with corrected oscillator strength values (log\,$gf$) for several lines, and using several molecular line lists : TiO (Plez 1998), 
VO (Alvarez \& Plez 1998), CN and CH (Hill et al., 2002). \\

We have progressively modified the stellar parameters of synthetic spectra until a good
agreement was achieved through fits to the observational spectra. Table 3 presents the stellar parameters and LTE lithium
abundances we derived using synthetic spectra displayed in Figure 1 around
the Li line region. The total error on the Li abundance has been estimated by computing the quadratic sum of errors 
on individual parameters. For HD 232\,862 and KU Peg, as several
observations were available, we have investigated the Li equivalent
widths for each individual spectrum and found them in very good
agreement, considering the S/N for each observation. Hence displayed
spectra in Fig.~1 refer to the observations taken on December 10, 2006
(HD 232\,862) and on September 05, 2007 (KU Peg).\\

\begin{table}
\caption{Stellar parameters and lithium abundances for our sample stars, as derived from 
synthetic spectra (see text). Error bars are indicated.  
}
\begin{flushleft}
\smallskip
\begin{tabular}{l l l c l l }
\noalign{\smallskip}
\hline
\hline
star  & date &     \Teff (K) & \logg & $[Fe/H]$ & \Ali  \\
  & dd/mm/yy & \scriptsize{$\pm$ 250 K}& \scriptsize{$\pm$ 0.5} & \scriptsize{$\pm$ 0.1 dex}& \scriptsize{$\pm$ 0.25 dex}\\
\hline

HD 232\,862  & 10/12/06  & 5000& 3.0 & -0.30& 2.45 \\
KU Peg  &  05/09/07   & 5000 & 3.0  &-0.15 &0.0 \\
HD 21\,018   &  05/09/07  & 5250 & 3.0 & 0.0 & 2.8  \\
\hline
\end{tabular}
   \end{flushleft}  
\end{table}

\begin{figure}
\psfig{figure=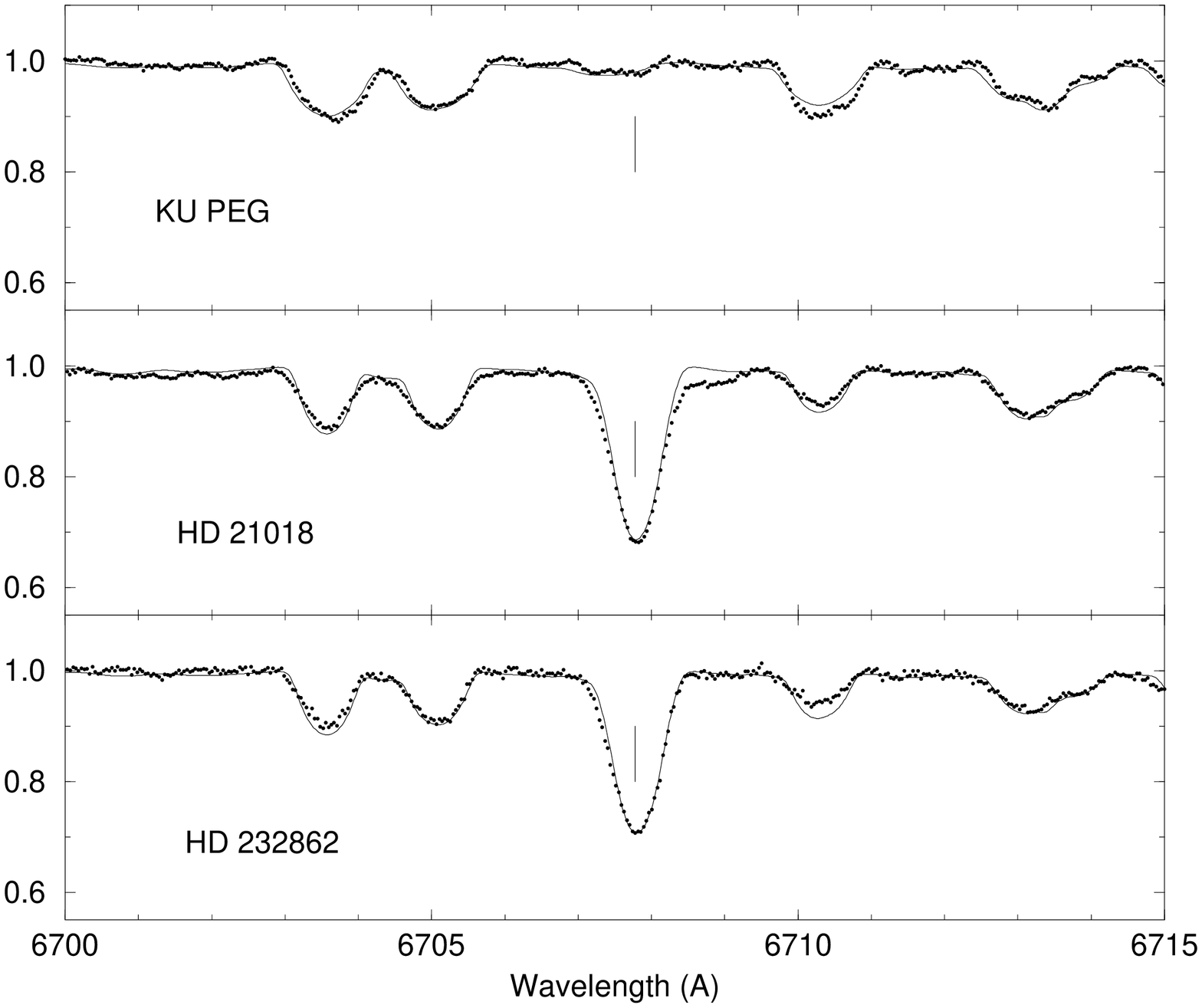,height=8.5cm,width=9.0cm,angle=270 }
\psfig{figure=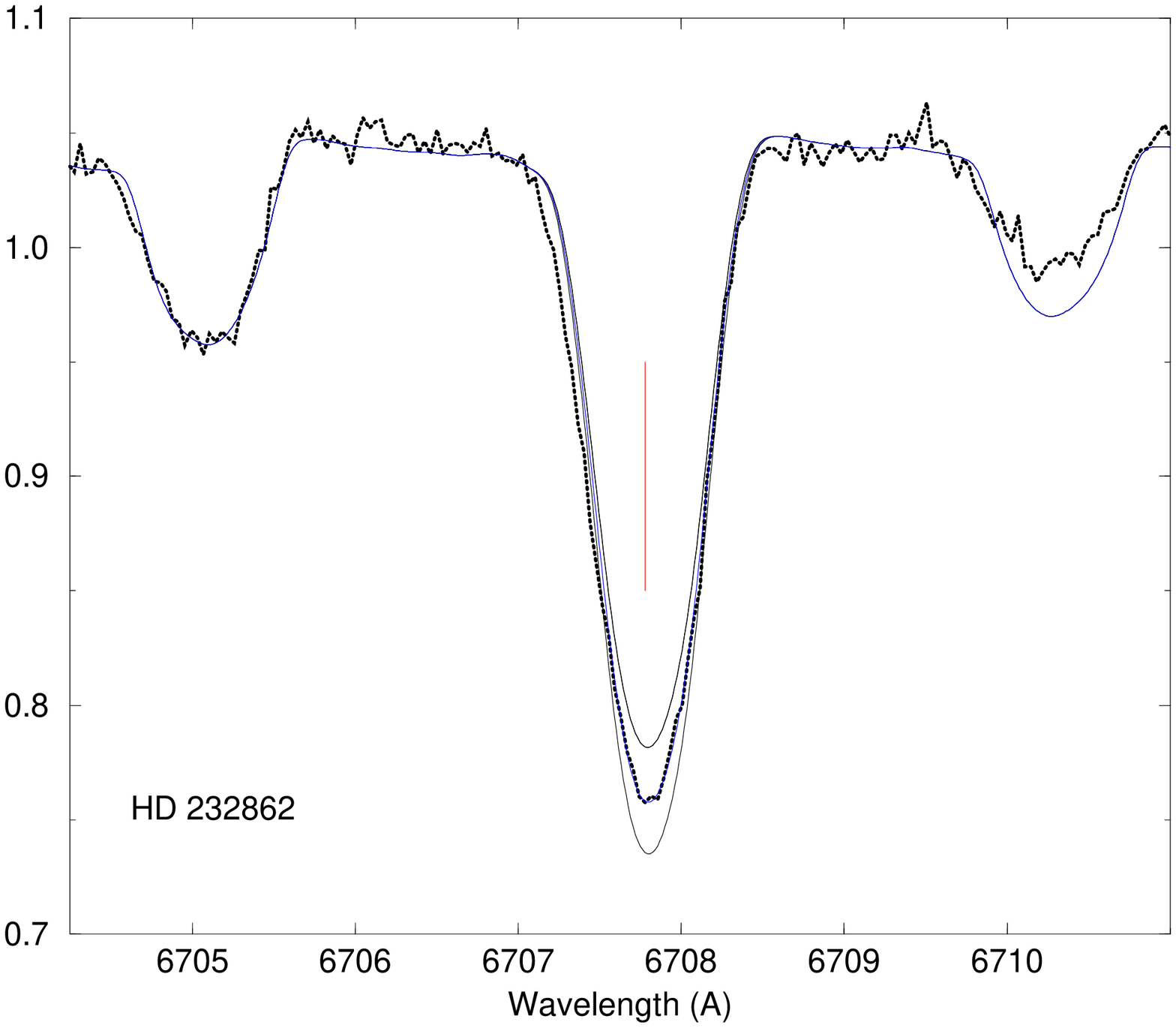,height=6.5cm,width=9.cm,angle=270 }
\caption[]{Spectral synthesis on our program stars (upper pannel) : observations (dotted line) in the Li line region (around 6707.8 \AA, 
indicated with the vertical tick line) and their associated synthetic spectra (full line) computed with the stellar 
parameters reported in Table~3, the rotational velocity values (see Table~1), and a microturbulence 
velocity sets to 2.0 \kms. \\
Lower panel : Zoom of the Li line in HD 232\,862 with three synthetic spectra computed with \Ali = 
2.35, 2.45 and 2.55 dex. Best fit is obtained with \Ali = 2.45 dex.\\
In both panels, the flux is normalized to the continuum.}
\end{figure}

Among our program stars, two objects present a high Li content : HD 21\,018, and our main target, 
HD 232\,862. As already reported by De Medeiros and L\`ebre (1992), the spectra of KU Peg show no sign of 
significant Li feature. \\

The rapidly rotating giant star HD 21\,018 exhibits a very high Li content : \Ali = 2.8 $\pm$ 0.25 dex. 
Previous spectroscopic studies devoted to HD 21\,018 have already mentioned this fact. Hartoog (1978) derived 
stellar parameters for HD 21\,018 in very good agreement with our determinations (\Teff = 5200 K and \logg = 3.0) 
and was the first to report its high Li content (\Ali = 2.93).  Later on, Barrado y Navascues et al. (1998) reported 
a much higher value for Li abundance (\Ali = 3.35 $\pm$ 0.4 dex), while adopting similar stellar parameters (\Teff = 5154 K). 
These two previous spectroscopic studies devoted to HD 21\,018 used a similar method (based on curves of growth) to derive 
Li abundance from a Li+Fe blended feature presenting the very same equivalent width (221-222 m\AA) in both works. 
Besides the use of spectral synthesis, our present Li abundance measurement in HD 21\,018 takes into account a 
better determination of \vsini and it remains - within respective error bars - rather compatible with previous determination. \\

But the surprising result here is the high Li content we derived for  HD 232\,862 : \Ali = 2.45 $\pm$ 0.25 dex. 
It is far in excess of the expected value for this spectral type (G8), and also quite peculiar for a bright giant.  
Indeed, for the relevant range of mass and evolutionary stage, surface lithium is expected to be diluted in  the convective envelope.\\

While there is no HIPPARCOS parallax available for HD 232\,862, a complementary chemical study focused 
on CNO products and $^{12}$C/$^{13}$C ratio would help to clarify its evolutionary 
status. Unfortunately, the rotational velocity of HD 232\,862 (Vsini = 20.6 km/s) is 
among the highest measured for this class of objects (De Medeiros \& Mayor, 1999). It turns out that it was not possible 
to assess its $^{12}$C/$^{13}$C isotopic ratio even from very good ESPaDOnS spectra (we vainly tried from features 
around 4200 \AA~and 8000 \AA).

\section{Surface magnetic field detections}

\subsection {Methodology}

As observations were collected with the spectropolarimeters ESPaDOnS and NARVAL (see Sect. 2), 
it was also possible to detect the presence of a magnetic field at the stellar surface. 
 Stokes V profiles enable to detect circular polarization on 
lines, which is related to the longitudinal Zeeman effect produced by the presence 
of a magnetic field in the stellar photosphere (Landstreet, 1992).\\ 

The extraction of polarization echelle spectra has been performed using the Libre-ESpRIT 
software (see Donati et al., 1997 for a full description of the original software package ESpRIT). 
To better study the Stokes V parameter profile, a Least Square Deconvolution (LSD) process was  
performed on the data right after each observation. The LSD tool (Donati et al., 1997, 
Shorlin et al., 2002) allows the extraction of mean circularly polarized and unpolarized profiles 
in a process similar to a cross correlation method. This technique - involving several thousands 
of spectral lines - offers the opportunity to considerably enhance the sensitivity of Zeeman detection, 
 when compared  to a single line analysis. When a Zeeman Stokes V signature is detected 
(cases of HD 232\,862 and KU Peg, see further), a null spectrum N (obtained by the standard procedure 
described in Donati et al., 1997) can be explored. It can help to confirm the detection of a clear 
Zeeman Stokes V signature when a feature is present only in V and not in N, and well located within 
the line profile velocity interval. Hence, the LSD analysis also provides a criterion for the detection 
of Zeeman Stokes V signatures and a detection probability: definite detection or marginal detection 
or no detection at all. When a definite detection is established, the longitudinal magnetic field 
$B_{l}$ is finally computed using the first order moment method (Donati et al., 1997, Rees \& Semel 1979). \\

\subsection {HD 232\,862 }

Four circularly polarized spectra  were recorded with ESPaDOnS on HD 232\,862, 
from December, 7, 2006 to December, 10, 2006 (see Tab. 2 in Sect. 2). To perform the LSD analysis, a specific 
mask involving a line list (with about 12 000 lines) computed from an ATLAS9 model atmosphere (from Kurucz, 2005) 
with \logg = 3.0 dex and \Teff = 5\,000 K has been used.\\

Figure 2 presents LSD profiles (December 7th, 2006 observation) and its definite magnetic field detection. 
The panel displays correlation profiles : Stokes I profile (lower plot, solid line), 
Stokes V profile (upper plot, dotted line) and the null spectrum diagnostic N (intermediate plot, dotted line). 
In order to help to distinguish the Zeeman signature from the noise, the V and N profiles have 
been vertically offseted (respectively by 1.1 and 1.0), expanded by a factor 25, and smoothed 
(solid lines). Figure 3 presents the Stokes V profiles of HD 232\,862 for the four consecutive 
observational dates (from December 7th to 10th, 2006). Magnetic field is thus detected unambiguously on each observations 
of HD 232\,862. The stokes V profile appears to be variable in time and 
also very complex, presenting several sign reversals throughout the line profile. 
The complex aspect and variability of the  detected magnetic signatures strongly suggest that the parent field 
structure may have a dynamo origin and a complex topology, likely with field topology variations on a rotational 
period that still needs to be determined. \\

\begin{figure}
\begin{center}
\psfig{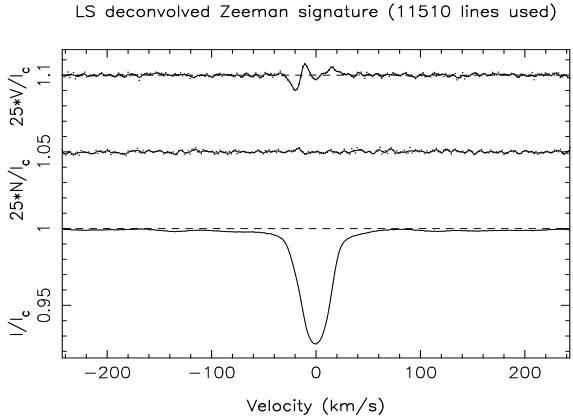}
\caption{HD 232\,862: Least Square Deconvolved Zeeman signature extracted from ESPaDOnS 
spectropolarimetric observation (December 07, 2006). A specific 
mask  ( $\sim$ 12\,000 lines and atmospheric parameters :  \Teff = 5000 K and \logg = 3.0) has been used (see text).}
\label{fig:LSDHD232862}
\end{center}
\end{figure}
                                                                                                           
\begin{figure}
\begin{center}
\psfig{figure=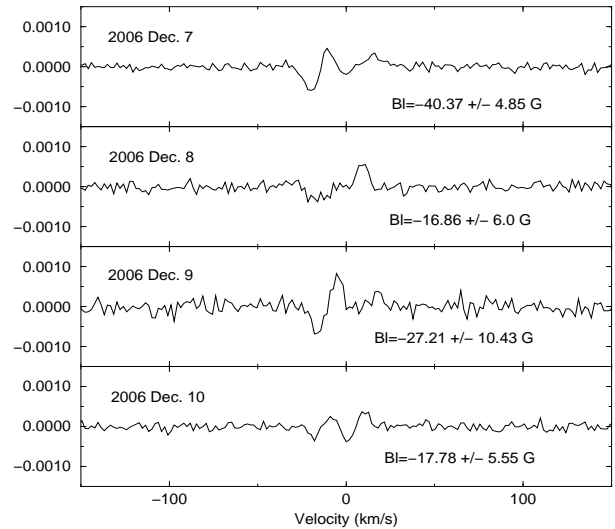,height=7.5cm,width=9.5cm,angle=270}
\caption[]{HD 232\,862: Time variation of Stokes V profiles obtained from the LSD method (with a specific 
mask composed of  about 12\,000 lines and computed with \Teff = 5000 K and \logg = 3.0), 
from December, 07, 2006 (upper plot) to December, 10, 2006 (lower plot).  
A clear signal, hosting a complex structure strongly variable from one night to another, 
is obtained in the Stokes V parameter. 
The longitudinal magnetic field component $B_{l}$ (in G) estimated on each individual observation 
is also indicated. }
\label{fig:tvarHD232862}
\end{center}
\end{figure}

  The $B_{l}$ values (and 1 $\sigma$ accuracy) 
are indicated on each Stokes V profiles of 
Figure 3 and summarized in Table 4. For HD 232\,862, 
the $B_{l}$ values are strongly variable along the four-days timescale.
This result is indeed one of the first direct field detection on an active late giant. 
Only very recently Konstantinova-Antova et al., (2008) and Auri\`ere et al., (2008) have reported direct detection of the presence and 
variability of a magnetic field in active single giants, using similar instrumentation (NARVAL at 
TBL) and technics (LSD analysis). They have also reported $B_{l}$ values variations. 
Magnetic field has also been detected from spectropolarimetric techniques 
in very fast rotators : a FK Com type giant (Petit et al., 2004) and in RS CVn type giants (Berdyugina et al.,
2006). The magnetic field in HD~232\,862 appears rather strong, but its complexity and rapid variations 
point to a dynamo-process origin in a rapid rotator rather than to the descendant of an Ap star, as in the case
of EK Eri (Auri\`ere et al., 2008).
More spectropolarimetric observations are needed in a 
monitoring mode, in order to ascertain the periodicity of the Zeeman Stokes V profiles variation, and to study the 
associated change in the field topology of HD~232\,862. 

\begin{table}
\caption{$B_{l}$ values (in G) for the magnetic field estimated on each individual observations of 
HD 232\,862 and KU Peg.}
\begin{flushleft}
\smallskip
\begin{tabular}{l l l l }
\noalign{\smallskip}
\hline
\hline
star  & date &  $B_{l}$    & $\sigma_{B_{l}}$ \\
  & dd/mm/yy &     &  \\

\hline

HD 232\,862 & 07/12/06 &  -40.37 & 4.85\\
HD 232\,862 & 08/12/06&  -16.86 & 6.00\\
HD 232\,862 & 09/12/06&  -27.21 & 10.43\\
HD 232\,862 & 10/12/06&  -17.78 & 5.55\\
KU Peg  & 04/09/07 & -3.15 & 4.96\\
KU Peg  & 05/09/07 & -9.02 & 4.44\\

\hline
\end{tabular}
   \end{flushleft}  
\end{table}

\subsection {KU Peg} 

Two circularly polarized spectra  were recorded with NARVAL at TBL on the active star KU Peg
on September, 4 and 5, 2007 (see Tab. 2 in Sect. 2). As on HD~232\,862 observations, a LSD analysis has 
been performed on these data, just after the exposure, each time concluding to definite magnetic field detections. 
The LSD quick look performed on the observation of the 2007 Sept. 5th is displayed in Figure 4.
Again, the specific mask (\Teff = 5\,000 K and  \logg = 3.0  dex) involved in the analysis of HD 232\,862 
was used to compute the correlation profiles. 
Figure 5 presents these two LSD 
profiles and their definite magnetic field detections. Variation on the Zeeman Stokes V profiles is also 
present in KU Peg data. The longitudinal magnetic field $B_{l}$ has also been estimated (see Tab.~4). 
It ranges from 
3 to 9 G, and appears, at this phase, weaker than the one estimated for HD 232\,862. This weak value
is consistent with the theoretical prediction by Kitchatinov \& R\"udiger (2004) for KU Peg. 
More spectropolarimetric observations are needed to confirm this first direct detection of a magnetic field 
in this chromospherically active giant, and to determine the intensity of its magnetic field. 

\begin{figure}
\begin{center}
\psfig{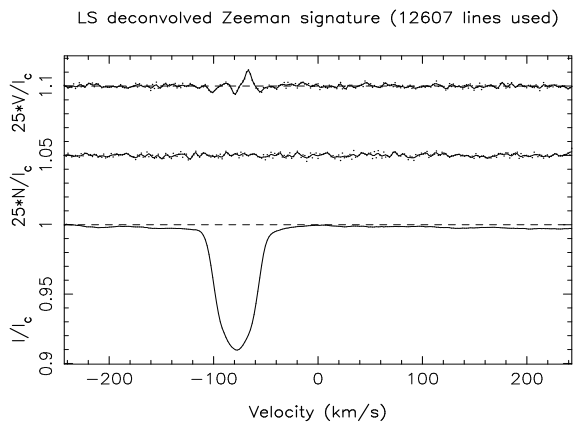}
\caption{KU Peg : Least Square Deconvolved Zeeman signature extracted from NARVAL  
spectropolarimetric observation (September 05, 2007). Same as Figure 2.}
\label{fig:LSDKUPEG}
\end{center}
\end{figure}

\begin{figure}
\begin{center}
\psfig{figure=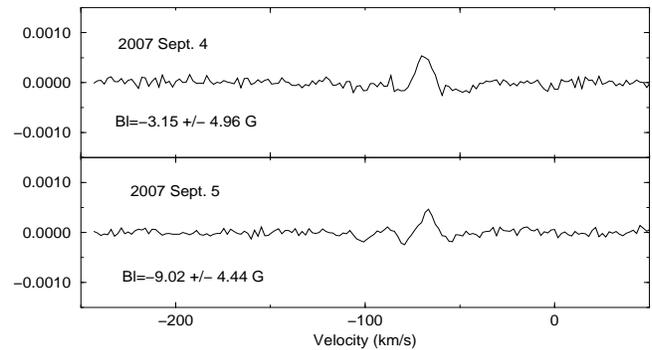,height=7.5cm,width=9.5cm,angle=270 }
\caption{KU Peg : Time variation of Stokes V profiles obtained from the LSD method (with a specific 
mask composed of  about 12\,000 lines and computed with \Teff = 5000 K and \logg = 3.0), 
from September, 04, 2007 (upper plot) to September, 05, 2007 (lower plot).  
A clear signal is obtained in the Stokes V parameter. 
The longitudinal magnetic field component $B_{l}$ (in G) estimated on each individual observation 
is also indicated. }
\label{fig:tvarVKUPEG}
\end{center}
\end{figure}

\subsection {HD 21\,018}

From the NARVAL observations of HD 21\,018 no Zeeman Stokes V
detection has been obtained (Figure 6).
Again, more
spectropolarimetric observations are needed to confirm the absence of
Zeeman signature and/or to exclude the presence of a very weak
field.\\

\begin{figure}
\begin{center}
\psfig{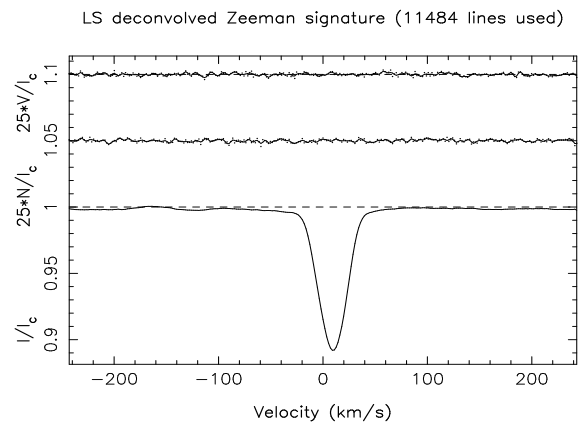}
\caption{HD 21\,018  : Least Square Deconvolved Zeeman profile performed on spectropolarimetric 
observation of September 05, 2007. A specific mask (composed of  about 12\,000 lines) and computed 
with \Teff = 5250 K and \logg = 3.0 has been used.}
\label{fig:LSDHD21018}
\end{center}
\end{figure}

\section{Evolutionary status}

In this section, we wish to address the origin of 
the high Li abundance in HD 232\,862 in terms of chemical evolution.
 The Li content we derive for this G8 star is indeed far in excess of the expected value 
for this spectral type, and also quite peculiar for a bright giant.
In the observational lithium survey of bright giants of L\`ebre et al., (2006), only few stars 
located on the blue side of the 
Hertzsprung gap, i.e. with \Teff $>$ 5500 K, show a high Li content (\Ali $>$ 2.0).  
For single and binary stars, almost all stars located on the red side of the Hertzsprung gap, 
ie. in the G to K spectral type range, show low Li content. 
However, for a lower mass range object ( class of luminosity of III : stellar mass $<$ 2.5 \Msun) 
this Li content, although exceptionally high, 
can be explained but at a very specific evolutionary point on the RGB lifetime : the so-called RGB bump.\\

We thus need to determine the evolutionary status of this star, and
more precisely, to evaluate its location with
  respect to the first dredge-up advancement. Unfortunately, due to
the lack of a precise parallax available for HD 232\,862 it is
difficult to assess precisely its evolutionary stage. Moreover its
high rotational velocity prevents a reliable measurement of its
$^{12}$C/$^{13}$C carbon isotopic ratio even from a high resolution
and good signal to noise spectrum collected with a 4m class
telescope.\\ 
In order to evaluate an approximate evolutionary
  status and mass for HD 232862, we have decided to directly compare its 
  derived stellar parameters and Li abundance presented in Table 3 
  with the predictions from standard single star stellar
  evolution models at the same metallicity (i.e. [Fe/H] = -0.3).\\
To
  do so, we have used the STAREVOL V3.0 stellar evolution code. A
  detailed description of STAREVOL V2.90 can be found in Siess (2006). Let us
  recall the main physical ingredients used for the models presented here.  
  We have considered a reference solar composition following Grevesse \& Sauval (1998), and we have 
  applied a simple scaling to derive the initial mixture for the different metallicities considered, 
  which is the usual approach when computing at non-solar metallicities (Girardi et al., 2000 ; Schaller et al. 1992). 
 In this case, at [Fe/H] = -0.3, the initial lithium abundance is \ALi = 2.976~dex. Mass loss was
  included from the Zero Age Main Sequence (ZAMS), following Reimers (1975)
  law with a parameter $\eta_{Reimers} = 1.00$. No transport of matter
  was considered in the radiative interiors, the only ``mixed''
  regions being the convective ones. Convection is modeled following
  the Mixing Length Theory, with a parameter $\alpha_{mlt} = L/H_p =
  1.75$. The Schwarzschild criterion was applied and no convective
  overshooting was considered.\\
  
\begin{figure}
\includegraphics[width=0.45\textwidth,height=7.5cm]{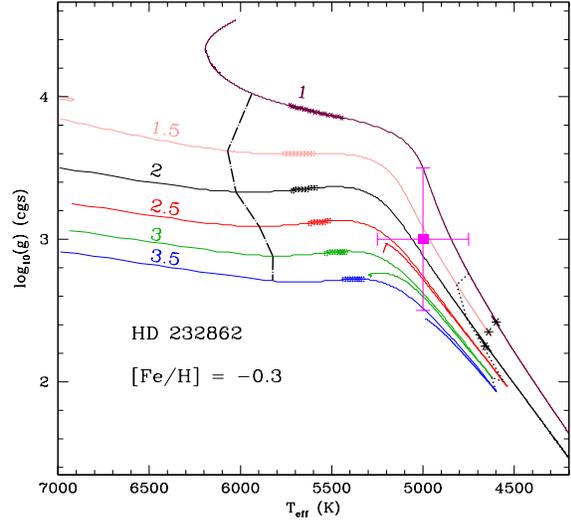}
\caption{Position of HD 232\,862 in the (\logg, \Teff) plane
    (filled square with attached error bars). The tracks have been
    computed for [Fe/H] = -0.3. The long-dashed-dotted line crossing
    through the tracks on the left indicates the beginning of the
    first dredge-up (DUP). The dotted line on the right indicates the
    location of the maximum penetration (in mass) of the convective
    envelope.  The bold parts on each track indicate the location where
    \ALi $\in [2.2; 2.7]$, corresponding to the interval deduced from spectroscopic 
    data for the lithium abundance of HD 232\,862 (see Table~1). The numbers refer 
    to the mass associated to
    each track (in solar masses).  The asterisks on the tracks for 1, 1.5 and 2 M$_\odot$ 
    indicate the location of the RGB bump on these tracks.}
\label{fig:gvsTHRD232862}
\end{figure}

We can get a first flavour of the evolutionary status of HD 232\,862
from its position in the (log g, T$_{eff}$) plane. In
Figure 7, we have placed the values derived for HD
232\,862 in a graph along with tracks for models of  1.0, 1.5,
2.0, 2.5, 3.0 and 3.5 M$_\odot$, covering the lower end of its
expected mass domain.  From the derived values of log g and \Teff
  quoted in Table~1, HD 232\,862 appears to be in the mass range
  between 1 M$_\odot$ and 3.5 M$_\odot$, with a prefered mass of about
  2 M$_\odot$. In this mass range, the models shown in
  Figure 7 indicate that the star is currently
  undergoing the first dredge-up, and hence lithium dilution. On the
tracks we have indicated by bold parts the 
 portion 
where the predicted Li surface abundance is within the
  determined errorbar, i.e. between 2.7 and 2.2 dex. The determined surface gravity and effective 
  temperature clearly place HD 232\,862 on the right side of these locations. As lithium abundance 
  decreases at the surface as the dredge-up proceeds, this means that according to standard stellar 
  evolution models, HD 232\,862 should have a much lower surface lithium abundance (between 0.9 and 1.9 dex) 
  than the high value of \ALi = 2.45 $\pm$ 0.25 dex estimated
  for this star from the spectroscopic analysis. In order to check this result, we have computed several other 
  grids, varying the initial metallicity of the models within the observed errorbar and assuming an initial lithium 
  abundance of 3.2 dex. In none of these cases are we able to reconcile the derived (\logg,\Teff) values with the 
  predicted lithium surface abundance.

  Thus, from these models, we might
  actually consider its high Li content to be particularly peculiar.  Indeed,
  a few field RGB stars present \ALi far in excess from the values
  predicted by standard theory. These are usually found at the
  so-called bump, an episode occurring higher on the RGB, after the
  completion of the first dredge-up, during which the surface Li
  abundance is reduced by a large factor (Charbonnel \& Balachandran
  2000). At this specific evolutionary point, the mean
  molecular-weight barrier left by the first dredge-up at its maximum
  penetration depth is erased by the outward moving hydrogen-burning
  shell (HBS). This may allow non-standard transport processes, such
  as rotation induced mixing, to connect freely the base of the
  convective envelope to the regions that have been processed by the
  HBS. Palacios et al.,  (2001) suggested production of fresh
  $^7$Li from $^7$Be at the bump, when the HBS first connects the
  $\mu$-barrier. This $\mu$-barrier is however very strong on the lower
  part of the RGB, where HD 232\,862 appears to lie, which should rule
  out this specific scenario\footnote{For the 1.0 M$_\odot$, 1.5 M$_\odot$ 
  and 2 M$_\odot$ models, the bump is located 
  around (L, T$_{eff}$, log g) = (43 L$_\odot$, 4580 K, 2.39), (75 L$_\odot$, 4640 K, 2.35) 
  and (120 L$_\odot$, 4670 K, 2.25) respectively, beyond the estimated location of HD 232\,862.}.\\

\begin{figure}
\includegraphics[width=0.45\textwidth,height=7.5cm]{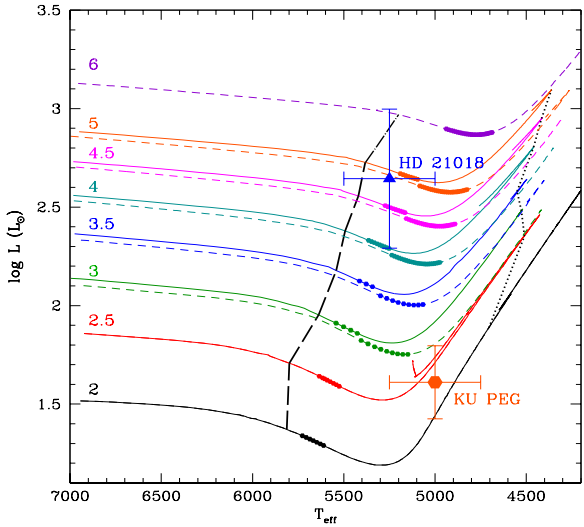}
\caption{Position of HD 21\,018 and KU~Peg in the
    Hertzsprung-Russell diagram. Solid and dashed tracks were computed
    at [Fe/H] = -0.15 (approximate metallicity of KU Peg) and [Fe/H] =
    0.0 (metallicity of HD 21\,018) respectively. The long-dashed line
    crossing through the tracks on the left indicates the beginning of
    the first DUP for the models at [Fe/H] = -0.15. The dotted line on
    the right indicates the location of maximum penetration (in mass)
    of the convective envelope for these same tracks. The small
    diamonds on the tracks indicate the location where  \ALi is within the errorbar derived for HD 21\,018, i.e. $\in [2.55;
      3.05]$}. The numbers refer to the mass associated to each
    track (in solar masses). 
\label{fig:HRKUPeg}
\end{figure}

\begin{figure}
\includegraphics[width=0.45\textwidth,height=7.5cm]{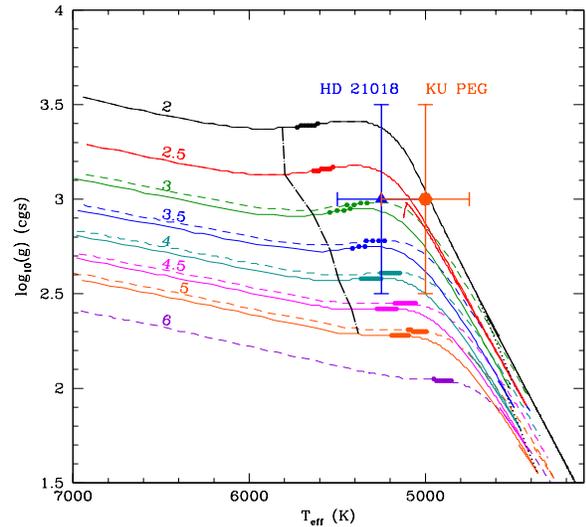}
\caption{Position of KU Peg and HD 21\,018 in the (\logg, \Teff) plane. 
See Figure 7 for detailed description of tracks and lines.}
\label{fig:gvsTKUPeg}
\end{figure}

In order to validate the approach we have adopted to estimate the
evolutionary status of HD 232\,862, we have applied the same method to
KU Peg and HD 21\,018, both stars for which Hipparcos parallaxes are
available  which allows us to derive their luminosity. We have
computed two grids of models between 2 M$_\odot$ and 6 M$_\odot$ at
[Fe/H] = -0.15 and 0.0, the metallicities associated to KU Peg and HD
21\,018 respectively. Figure~8 presents the Hertzsprung-Russell
diagram for these grids. In order to locate our two stars in this
theoretical plane, we need to transform the observed magnitudes into
luminosities. We have applied the bolometric corrections from Bessell
et al., (1998), taking into account the temperature and surface
gravity of each star. The adopted values are BC$_V$ = -0.26 and BC$_V$
= -0.17 for KU Peg and HD 21\,018 respectively. The interstellar
absorption A$_V$ is given by the reddening maps available via the
NASA/IPAC Infrared Science Archive. The adopted values are $A_V$ =
0.3205 and 0.4274 for KU Peg and HD 21\,018 respectively.\\ As in
Fig.~7, we have indicated the location of the beginning and of the end
of the first dredge-up, as well as the range of temperature and
luminosity where A$_{Li}$  varies between 3.05 and 2.55 dex. KU
Peg appears to have a mass between 2 and 2.5 M$_\odot$, where the Li
dilution is already well advanced, which is in agreement with no clear
signature of Li found in this star (De Medeiros \& L\`ebre, 1992).  In
good agreement with Charbonnel \& Balachandran (2000) and Denissenkov
et al., (2006), we estimate the mass of HD 21\,018 to about 5
M$_\odot$. According to Fig.~7, the first dredge-up episode has just
started in HD 21\,018, and its Li surface abundance should not be
altered yet, in agreement with the large abundance determined for this
star.\\ Let's turn now to the position of these stars in the (\logg,
\Teff) plane, displayed in Figure 9.
The location of
KU Peg is in perfect agreement with its position on the HR-diagram,
and we confirm its low mass and more advanced evolutionary
status. Concerning HD 21\,018, the agreement is not as good, and the
cross comparison of Figs.~7 and~8 points to a smaller mass, likely
around 4 M$_\odot$. The rather large uncertainty on the Hipparcos parallax 
is in part responsible of this marginal agreement between the HR diagram and the (\logg,\Teff) diagram. 
This star should however be at the beginning of
the first dredge-up episode, in agreement with observations.

\section {Conclusions}
We have determined a high Li content (\ALi = 2.45 dex $\pm$ 0.25 dex) in the field star HD~232\,862,  classified as a 
bright giant (luminosity class II -- stellar mass $>$ 2.5 \Msun), and this fact could constitute 
the first report of a high Li-content in a bright giant star. 
However, considering its evolutionary status, HD~232\,862 appears to  be in the lower part of the mass 
interval [1.0 M$_\odot$, 3.5 M$_\odot$], likely a 1.5  or 2 M$_\odot$, located at 
the bottom of the Red Giant Branch. A mis-classification of HD~232\,862 among the luminosity class II objects is then suggested. 
Anyway, the high Li-content we have found in this rapidly rotating giant is very peculiar. 
Indeed, from standard stellar evolution the star should have already experienced deep enough dredge-up so as to 
exhibit low lithium content at its surface. On another hand, HD~232\,862 cannot be considered as one of those Li-rich giants located 
at the RGB bump as, from its location in the (log g, T$_{\rm eff}$) plane, the star would not have reached this specific evolutionary 
stage yet.\\

 Hence, the high Li content we report in HD 232\,862 is still puzzling. We can not rule out that it is due to a 
mecanism preventing the  efficient Li dilution  expected during the 1rst dredge--up to occur.\\

Moreover, we have detected a surface magnetic field on HD 232\,862. It strongly varies during 4 days observations, pointing 
likely to a dynamo nature. Rough estimate of its longitudinal component reveals a high intensity. To date this magnetic 
field we have detected in HD~232\,862 seems to be one of the  strongest field 
ever detected in a giant, and also the very first one to be detected in a Li-rich giant. Further observations (and modeling of 
V profiles) are now needed to infer the mean magnetic field of HD~232\,862, to clarify its origin, and to assess the progenitor of 
HD~232\,862. Only then, will it be possible to disentangle whether its high \ALi is due to the 
presence of a surface magnetic field, to the binary status of HD~232\,862, or to another process.  \\

The magnetic field of the rapidly rotating active giant star KU Peg has also been detected for the first time. It appears 
to be weaker than the one measured in HD~232\,862. More spectropolarimetric observations, in a monitoring mode along the rotational 
period, are now needed to better study the topology of this magnetic field. Such spectropolarimetric observations of KU Peg and HD 232\,862 
are already in progress with NARVAL instrument.

\acknowledgements

{AL thanks J.F. Donati for his kind help during the observing run at CFHT with the ESPaDOnS instrument, and 
C. Catala for his helpful supports during time sharing runs.
This research has made use of the following soft and services : the SIMBAD database, operated at the CDS 
(Strasbourg, France), the NASA's Astrophysics Data System, the NASA/IPAC Infrared Science Archive, which is 
operated by the Jet Propulsion Laboratory, California Institute of Technology, under contract with the National 
Aeronautics and Space Administration, and the Vienna Atomic Line Database (VALD). 
Spectropolarimetric data have been reduced and analysed, 
 respectively with the Libre-ESpRIT software and the Least Square Deconvolution routine, both written 
 by J.F. Donati (LATT, France) and kindly provided during the ESPaDOnS/CFHT and NARVAL/TBL observing runs. 
 We thank the french PNPS/INSU-CNRS for financial support and JDNJr. acknowledges the CNPq for a partial support.}


\begin{thebibliography}{}

\bibitem[1967]{Alexander67}
Alexander J.B., 1967,  Observatory 87, 238

\bibitem[1998]{Alvarez98}
Alvarez, R., \& Plez, B. 1998, A\&A, 330, 1109

\bibitem[1995]{Antov95}
Antov A.P., Konstantinova-Antova R.K., 1995, in Robotic Observatories, 
Bode M.F. (ed.), Praxis Publishing, Chichester, p.69 

\bibitem[2003]{auriere03}
Auri\`ere M., 2003, in Magnetism and Activity of the Sun and Stars, ed. J. 
Arnaud \& N. Meunier, EAS Publ. Ser., 9, 105

\bibitem[2008]{auriere08}
Auri\`ere M., Konstantinova-Antova R., Petit P. et al., 2008, A\&A 491, 499

\bibitem[1998]{barrado98}
Barrado y Navascues, D., de Castro, E., Fernandez-Figueroa, M.J., Cornide, M., Garcia Lopez, R.J.,  1998, A\&A 337, 739

\bibitem[1989]{Bell89}
Bell R.A., Gustafsson B., 1989 MNRAS 236, 653

\bibitem[2006]{Berdyugina06}
Berdyugina S.V., Petit P., Fluri D.M., et al., 2006, ASPC 358, 381B

\bibitem[1998]{Bessel98}
Bessell M.S., Castelli F., Plez B., 1998 A\&A 333, 231

\bibitem[1989]{brown89}
Brown J.A, Sneden C., Lambert D.L., Dutchover E., 1989, ApJS 71, 293

\bibitem[1971]{Cameron71}
Cameron A.G.W. \& Fowler W.A., 1971, \apj, 164, 111

\bibitem[2006]{CantoMartins06}
Canto Martins B.-L., L\`ebre A., De Laverny P. et al., 2006 A\&A 451, 993

\bibitem[1999]{Castilho99}
Castilho B.V., Spite F., Barbuy B., Spite M., de Medeiros J.R., Gregorio-Hetem J., 1999 A\&A 345, 249

\bibitem[2000]{Corinne00}
Charbonnel C. \&  Balachandran S.C., 2000, A\&A 359,563

\bibitem[1988]{Couteau88}
Couteau P., 1988 A\&AS 75, 163

\bibitem[1996]{delareza96}
de la Reza R., Drake N.A., Silva L., 1996, \apj~ 456, L115

\bibitem[2006]{delareza06} de La Reza, R.\ 2006, 
Chemical Abundances and Mixing in Stars in the Milky Way and its 
Satellites, ESO ASTROPHYSICS SYMPOSIA.~ISBN 
978-3-540-34135-2.~Springer-Verlag, 2006, p.~196 

\bibitem[2003]{delaverny03}
de Laverny P., do Nascimento J.D.Jr., L\`ebre A., De Medeiros J.R., 2003, A\&A 410, 937

\bibitem[1992]{jose92}
De Medeiros J.R., L\`ebre A., 1992, A\&A 264, L21

\bibitem[1992]{renan92}
De Medeiros J.R., Mayor M., Simon T., 1992, A\&A 254, L36

\bibitem[1996]{renan96a}
De Medeiros J.R., Da Rocha C., Mayor M., 1996, A\&A 314, 499 

\bibitem[1996]{renan96b}
De Medeiros J.R., Melo C.H.F., Mayor M., 1996, A\&A 309, 465

\bibitem[1997]{jose97}
De Medeiros J.R., do Nascimento J.D.Jr., Mayor M., 1997, A\&A 317, 701

\bibitem[1999]{jose99}
De Medeiros J.R., Mayor M., 1999, A\&AS 139, 443

\bibitem[2000]{denissenkov00}
Denissenkov P.A. \& Weiss A. 2000, A\&A 358, L49 

\bibitem[2006]{denissenkov06}
Denissenkov P.A., Chaboyer B., Li K., 2006, ApJ 641, 1087

\bibitem[1997]{donati97}
Donati J.-F., Semel M., Carter B.D., Rees D.E., Cameron A.C. 1997, MNRAS 291, 658 

\bibitem[2004]{donati04}
Donati J.-F., 2004, in "Scientific highlights 2004", EDP Sciences Conference Series, 
Eds. F. Combes et al., p.217

\bibitem[2006]{donati06}
Donati J.-F., Catala C., Landstreet J.D., Petit P., 2006, in "Solar Polarization 4", 
ASP Conf. Series, Vol.358, Eds. R. Casini \& B.W. Lites, p.362

\bibitem[2002]{drake02}
Drake, N. A., de La Reza, R., da Silva, L., Lambert, D.L., 2002, AJ
123, 2703

\bibitem[1993]{feke93}
Fekel F.C., Balachandran S., 1993, ApJ 403, 708

\bibitem[1996]{feke96}
Fekel F.C., Webb R.A., White R.J., Zuckerman B., 1996, ApJ 403, 708

\bibitem[1996]{Flower96}
Flower P.J., 1996, ApJ 469, 355

\bibitem[1997]{Gilli97}
Gili R., Couteau P., 1997 A\&AS 126, 1

\bibitem[2000]{Gir00}
Girardi L., Bressan A., Bertelli G., Chiosi C., 2000, A\&AS 141, 371

\bibitem[89]{gratton89}
Gratton, R.G., D'Antona, F. 1989, A\&A,  215, 66

\bibitem[1998]{Grevesse98}
Grevesse, N., \& Sauval, A. J. 1998, SSRv, 85, 161

\bibitem[2008]{Gustaffson2008}
Gustafsson, B., Edvardsson, B., Eriksson, K., Jorgensen U.G., Nordlund A., Plez B., 2008, 
A\&A 486, 951

\bibitem[1978]{Hartoog78} Hartoog, M.~R.\ 1978, \pasp, 90, 167 

\bibitem[1990]{Heintz90}
Heintz W.D., 1990 ApJS 74, 275

\bibitem[1998]{Heintz98}
Heintz W.D., 1998 ApJS 117, 587

\bibitem[2002]{Hill2002}
Hill, V., Plez, B., Cayrel, R., et al. 2002, A\&A, 387, 560
,
\bibitem[2000]{Hill2000}
Hill, V., Pasquini L., 2000, in "The Light Elements and their Evolution", IAU Symp. No~198,  
eds. L. da silva, R. de Medeiros, \& M. Spite, p.293

\bibitem[1965]{Iben65} 
Iben I.J., 1965, ApJ 142, 1447

\bibitem[1966]{Iben66a} 
Iben I.J., 1966a, ApJ 143, 483

\bibitem[1966]{Iben66b} 
Iben I.J., 1966b, ApJ 143, 505

\bibitem[1991]{Iben91} 
Iben, I.~J.\ 1991, \apjs, 76, 55 

\bibitem[2004]{kitch04}
Kitchatinov L.L., R\"udiger G., 2004, Astron. Nachr. 325, 496

\bibitem[2008]{Renada08}
Konstantinova-Antova R., Auri\`ere M., Iliev I.Kh., Cabanac R., Donati J.-F., 
Mouillet D., Petit P., 2008, A\&A, 480, 475

\bibitem[1999]{kraft99}
Kraft R.P., Peterson R. C., Guhathakurta P., Sneden C.,  Fulbright J.P.,
 Langer G.E.  1999, \apj 518, 53
 
\bibitem[1999]{kupka99}
Kupka F., Piskunov N.E., Ryabchikova T.A., Stempels H.C., Weiss W.W., 1999, A\&AS 138, 119

\bibitem[2005]{kurucz05}
Kurucz R., 2005, Mem. S.A. I. Suppl., Vol. 8, p~14

\bibitem[1992]{landstreet 92}
Landstreet J.D., 1992, AARev 4, 35

\bibitem[1999]{lebre99}
L\`ebre A., De Laverny P., De Medeiros J.R., Charbonnel C., Da Silva L.,
1999, A\&A 345, 936

\bibitem[2006]{lebre06}
L\`ebre A., de Laverny P., do Nascimento J.D., de Medeiros J.R.,
  2006, A\&A 450, 1173

\bibitem[1995]{Luck95}
Luck R.E., Wepfer G.G., 1995, AJ 110, 2425

\bibitem[2005]{Melo2005} 
Melo C.H.F., de Laverny P., Santos N.C., et al., 2005, A\&A, 439, 227

\bibitem[2008]{MonacoBonifacio2008} 
Monaco L., \& Bonifacio P.\ 2008, Memorie della Societa Astronomica Italiana, 79, 524 

\bibitem[2001]{Palacios01}
Palacios A., Charbonnel C., Forestini M., 2001 A\&A 375, L9

\bibitem[2003]{Palacios03}
Palacios A., Talon S., Charbonnel C., Forestini M., 2003 A\&A 399, 603

\bibitem[2004]{Petit04}
Petit P., Donati J.F., Oliveira J.M.,  et al., 2004 MNRAS 351, 826 

\bibitem[2000]{Pilachowski2000} 
Pilachowski C.~A., Sneden C., Kraft R.~P., Harmer D., \& Willmarth D.\ 2000, \aj, 119, 2895 

\bibitem[1998]{Plez1998}
Plez B. 1998, A\&AS, 337, 495

\bibitem[1999]{randich99}
Randich S., Gratton R., Pallavicini R., Pasquini L., Caretta E., 1999 A\&A 348, 487

\bibitem[2002]{reddy02}
Reddy B.E., Lambert D.L., Hrivnak B.J., Bakker E.J., 2002, \aj~ 123, 1993

\bibitem[2005]{ReddyLambert05} Reddy, B.~E., \& Lambert, D.~L.\ 2005, \aj, 129, 2831 

\bibitem[1979]{rees79}
Rees D.E., Semel M., 1979 A\&A 74, 1

\bibitem[1975]{reimers75}
Reimers D., 1975, Mem. Soc. R. Sci. Li\`ege, 8, 369

\bibitem[2008]{RoedererFrebeletal2008}
Roederer I.~U., Frebel A., Shetrone M.D., Allende Prieto C., Rhee J., Gallino R., Bisterzo S., Sneden C., Beers T.C., Cowan J.J., 2008 
\apj 679, 1549 

\bibitem[1999]{ryab99}
Ryabchikova T.A., Piskunov N.E., Stempels H.C., Kupka F., Weiss W.W., 1999, in Proc. of the 6th International 
    Colloquium on Atomic Spectra and Oscillator Strengths, Victoria BC, Canada, Physica Scripta T.83, p.162

\bibitem[1999]{Sackmann99}
Sackmann I-J., Boothroyd A.I., 1999, \apj~  510, 217

\bibitem[1992]{Schaller92}
Schaller G., Schaerer D., Meynet G., Maeder A., 1992, A\&AS 96, 269

\bibitem[2002]{Shorlin02}
Shorlin S.L.S., Wade G.A., Donati J.-F., Landstreet J.D., Petit P., Sigut T.A.A., Strasser S., 2002, A\&A 392, 637

\bibitem[1999]{Siess99}
Siess L., Livio M., 1999, \mnras~ 304, 925

\bibitem[2006]{Siess06}
Siess L., 2006, A\&A 448, 717

\bibitem[1988]{Strassmeier88}
Strassmeier K.G., Hall D.S., 1988, ApJS 67, 453

\bibitem[1982]{wall82}
Wallerstein G., Sneden C., 1982, \apj~  255, 577

\bibitem[2005]{Weber05}
Weber M., Strassmeier K.G., Washuettl A., 2005, Astron. Nachr. 326, 287 

\end{thebibliography}
\end{document}